\documentclass[aps,amsfonts,pra,twocolumn,showpacs]{revtex4-1}

\usepackage{subfigure}
\usepackage{textcomp}
\usepackage{graphicx}
\usepackage{amssymb}
\usepackage{amsmath}
\usepackage{bm}
\usepackage{amsmath, amsthm, amssymb}
\usepackage{dsfont}
\usepackage[colorlinks=true,linkcolor=blue,urlcolor=blue,citecolor=blue]{hyperref}
\usepackage{multirow}
\usepackage{url}

\def\beq{\begin{equation}}
\def\eeq{\end{equation}}
\def\bea{\begin{eqnarray}}
\def\eea{\end{eqnarray}}

\usepackage[usenames, dvipsnames]{color}

\begin{document}

\title{Kinetic theory of spin diffusion and superdiffusion in XXZ spin chains}

\author{Sarang Gopalakrishnan$^1$ and Romain Vasseur$^2$}
\affiliation{$^1$ Department of Physics and Astronomy, CUNY College of Staten Island, Staten Island, NY 10314;  Physics Program and Initiative for the Theoretical Sciences, The Graduate Center, CUNY, New York, NY 10016, USA}
\affiliation{$^2$ Department of Physics, University of Massachusetts, Amherst, MA 01003, USA}

\begin{abstract}

We address the nature of spin transport in the integrable XXZ spin chain, focusing on the isotropic Heisenberg limit. We calculate the diffusion constant using a kinetic picture based on generalized hydrodynamics combined with Gaussian fluctuations: we find that it diverges, and show that a self-consistent treatment of this divergence gives superdiffusion, with an effective time-dependent diffusion constant that scales as $D(t) \sim t^{1/3}$. This exponent had previously been observed in large-scale numerical simulations, but had not been theoretically explained. We briefly discuss XXZ models with easy-axis anisotropy $\Delta > 1$. Our method gives closed-form expressions for the diffusion constant $D$ in the infinite-temperature limit for all $\Delta > 1$. We find that $D$ saturates at large anisotropy, and diverges as the Heisenberg limit is approached, as $D \sim (\Delta - 1)^{-1/2}$.

\end{abstract}

\maketitle

Integrable models play a central role in our understanding of quantum dynamics. These models not only allow exact calculations of otherwise intractable aspects of many-body dynamics, but also exhibit distinctive phenomena, such as their failure to thermalize from generic initial conditions~\cite{PhysRevLett.98.050405,Rigol:2008kq}. Nonequilibrium dynamics, transport, and entanglement in integrable and nearly integrable models have been topics of considerable recent interest, from both theoretical~\cite{Calabrese:2006, PhysRevLett.106.217206, PhysRevLett.110.257203, PhysRevLett.113.117202, PhysRevLett.115.157201, 2016arXiv160300440I, 1742-5468-2016-6-064001,1742-5468-2016-6-064002,1742-5468-2016-6-064010,1742-5468-2016-6-064007,PhysRevB.89.125101, alba2017entanglement} and experimental~\cite{kinoshita, gring, tang2018, erne2018, zundel2018} points of view.

There are two complementary ways of thinking about one-dimensional quantum integrable systems: these systems have stable, ballistically propagating quasiparticles; they also have a complete set of local and quasilocal conserved charges~\cite{PhysRevLett.106.217206, PhysRevLett.110.257203, PhysRevLett.113.117202, PhysRevLett.115.157201}, which can be related to the moments of the quasiparticle distribution. The presence of ballistic quasiparticles might suggest that transport of the conserved charges should be ballistic, even at high temperature; however, this is not always the case. In many systems, such as the isotropic and easy-axis XXZ spin chains, the Drude weight for certain charges (in this case, magnetization) vanishes~\cite{PhysRevB.79.214409,PhysRevB.80.184402,1742-5468-2009-02-P02035,PhysRevLett.106.220601, PhysRevLett.107.250602}. In these cases, the ballistic motion of quasiparticles (and thus of energy and quantum information) coexists with sub-ballistic spin transport. Depending on the parameters, spin transport can be either diffusive or superdiffusive~\cite{PhysRevLett.106.220601, lzp, idmp, sanchez2018anomalous}. 

Direct calculations of transport in interacting integrable models are challenging; however, the theory of generalized hydrodynamics (GHD)~\cite{Doyon, Fagotti} has emerged as a description of the long-wavelength, long-time dynamics of these systems~\cite{Doyon, Fagotti,SciPostPhys.2.2.014,PhysRevLett.119.020602, BBH0, BBH,PhysRevLett.119.020602, GHDII, doyon2017dynamics, solitongases,PhysRevLett.119.195301,PhysRevB.97.081111, dbd1, ghkv}. The main assumption of GHD is that the system is locally in a generalized Gibbs ensemble, with parameters that vary smoothly in space. This reduces the problem of computing dynamics in integrable systems to the considerably simpler one of computing \emph{thermodynamics} in these systems. GHD has been successfully used to compute Drude weights in integrable models~\cite{PhysRevLett.119.020602, BBH,GHDII,PhysRevB.96.081118,PhysRevB.97.081111}, and was recently generalized to account for Gaussian fluctuation effects~\cite{dbd1, ghkv}, which give rise to diffusive corrections~\cite{percus, sachdev_young, PhysRevLett.78.943, PhysRevB.57.8307, el2003thermodynamic, el2005, PhysRevLett.95.187201, sirker:2010, PhysRevB.83.035115, medenjak2017, kormos2017,  piroli2017, klobas2018a, klobas2018b, sg_ffa, dbd1, ghkv} to ballistic quasiparticle motion. However, these diffusive corrections suggest no obvious mechanism for \emph{superdiffusion}, which occurs in the isotropic XXX spin chain~\cite{lzp, idmp}. 


The present work offers a self-consistent theory of superdiffusion in the isotropic limit of the XXZ spin-$\frac{1}{2}$ chain. The general XXZ spin chain is described by the Hamiltonian
\beq\label{eq1}
H =  \sum\nolimits_i S^x_i S^x_{i+1} + S^y_i S^y_{i+1} + \Delta S^z_i S^z_{i+1}.
\eeq
We primarily consider the infinite temperature limit, and work at half-filling. In this limit, transport coefficients are, strictly speaking, zero; however, autocorrelation functions remain well-defined, and one can classify the high-temperature limit of transport based on their asymptotics. In the XXZ model, energy transport is purely ballistic regardless of $\Delta$, as the energy current is conserved under the dynamics. Spin transport, however, depends much more nontrivially on $\Delta$~\footnote{We are concerned here with finite-temperature states but \emph{not} with generalized Gibbs ensembles with finite magnetization. In the finite-magnetization sectors, spin transport remains ballistic for all $\Delta$.}. In the easy-plane regime $\Delta < 1$, spin transport has a ballistic component, with a Drude weight that varies nontrivially with $\Delta$~\cite{PhysRevLett.106.217206, PhysRevLett.119.020602}. When $\Delta \geq 1$, the spin Drude weight vanishes, so spin transport must be sub-ballistic. In the easy-axis regime $\Delta > 1$, spin transport is believed to be diffusive~\cite{PhysRevLett.106.220601,PhysRevB.89.075139,1367-2630-19-3-033027, idmp}. Thus an unusual high-temperature dynamical phase transition takes place in the XXZ spin chain, between an easy-plane ``phase'' with a nonzero Drude weight, and an easy-axis ``phase'' with zero Drude weight and diffusive transport. The mechanism for this phase transition is a qualitative change in the character of the conserved charges as one crosses the isotropic point.

One might expect the diffusion constant to diverge as one approaches $\Delta = 1$ from the easy-axis side, as indeed it does: a rigorous lower bound can be derived for the spin diffusion constant, and this lower bound diverges as $\Delta \rightarrow 1$, indicating that transport at the isotropic point must be superdiffusive~\cite{idmp}. The nature of transport at the isotropic point has not yet been explained theoretically, however. The numerical evidence on \emph{high-temperature} spin transport indicates that spin diffusion is anomalous; small spin imbalances appear to spread with the length-time scaling $x \sim t^{2/3}$~\cite{lzp}. Note that superdiffusion in this translation-invariant system must involve conceptually different mechanisms from phenomena such as Levy flights, which typically rely on disorder~\cite{bouchaud1990}. Indeed, the numerical evidence~\cite{lzp} suggests diffusive behavior but with a rescaled time coordinate: {\it i.e.}, the shape of the magnetization front is an error function in the appropriately scaled variables. 

Our main result here is a derivation of this exponent $2/3$ using ideas from GHD and its Gaussian corrections. We analyze the spreading of an initially localized ``packet'' of spin density. The packet spreads through the ballistic motion of quasiparticles, which (for $\Delta \geq 1$) are an infinite family of ``strings'' (magnons and boundstates thereof) parameterized by a variable $s$ called the string length~\cite{Takahashi}. When $\Delta > 1$, spreading is dominated by short strings because the longer ones are immobile (as explained below). As $\Delta \rightarrow 1$, strings at all scales begin to contribute to transport, as first noted in Ref.~\cite{idmp}. At the isotropic point, on a given timescale $t$ the dominant contribution is from the longest strings that are mobile on that timescale. This leads naturally to a time-dependent diffusion constant, and thus to superdiffusion.

\begin{figure}[tb]
\begin{center}
\includegraphics[width = 0.45\textwidth]{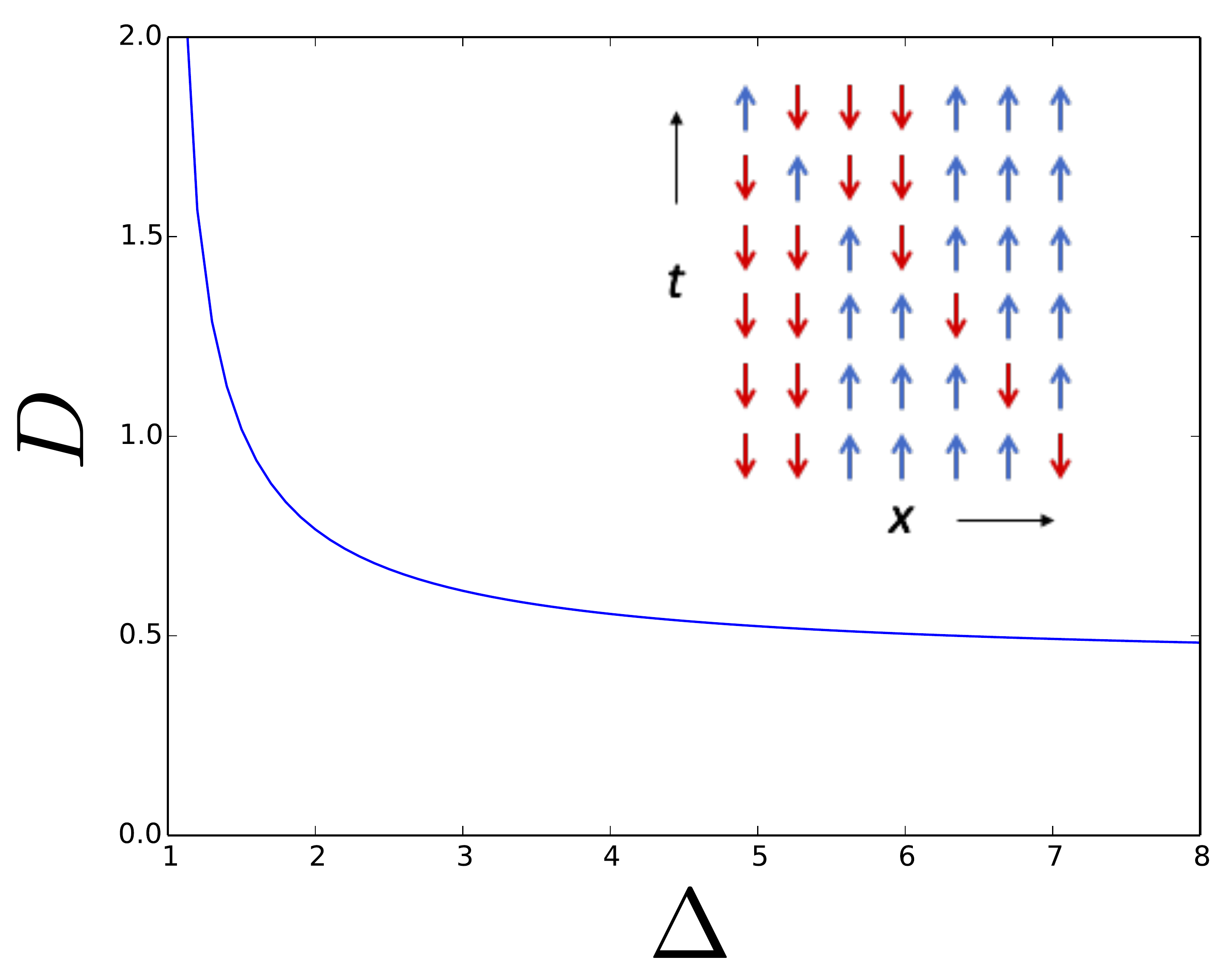}
\caption{Spin diffusion constant $D$ in the easy-axis phase $\Delta >1$ as a function of $\Delta$. The diffusion constant diverges in the anisotropic limit, leading to superdiffusion with $D(t) \sim t^{1/3} $. {\it Inset:} Propagation of a magnon in the large $\left| \Delta \right|$, low temperature, ferromagnetic limit. The magnon is initially a minority $\downarrow$ spin in an $\uparrow$-spin domain; it is then transmitted into a $\downarrow$-spin domain as a minority $\uparrow$ spin. On average, therefore, the magnon is neutral at half-filling, giving rise to a vanishing spin Drude weight. However, its effective magnetization fluctuates, leading to diffusive spin transport coexisting with ballistic spreading of energy and, presumably, quantum information.}
\label{FigD}
\end{center}
\end{figure}

\emph{Dressed magnetization fluctuations}.---The basic mechanism for diffusion in the easy-axis regime is easiest to describe for $\left| \Delta \right| \gg 1$ at relatively low temperatures in the ferromagnetic regime for the Hamiltonian~\eqref{eq1}. (However, the result itself applies generally, as we will see below.) Consider a snapshot of a typical thermal state; it will consist of alternating $\uparrow$-spin and $\downarrow$-spin domains of varying sizes. Most of these domains are immobile on short timescales at large $\Delta$, as the only energy- and spin-conserving processes are those in which the entire domain collectively moves, and these occur at high orders in perturbation theory. Dynamics is dominated instead by rare short domains, for instance a single $\uparrow$ spin in a large $\downarrow$-spin domain. The $\uparrow$-spin can propagate freely within the domain; it then propagates into the neighboring domain through a process where it is converted into a $\downarrow$ spin~\cite{quantumbowling}. As this excitation (which is a magnon) propagates ballistically through the sample, it spends half its time as an $\uparrow$ spin and half as a $\downarrow$ spin; on average, therefore, the magnon carries no magnetization. Thus it contributes to ballistic energy transport but not ballistic spin transport. (See inset of Fig.~\ref{FigD}.)

We now give an argument for spin transport being diffusive in this model~\cite{mkp, idmp}. We first present the argument in a less general but more elementary form. 
Consider an excess of spin initially localized at the origin; to compute the spreading of the spin at later times, we track the mean-squared ``dipole moment'' $\langle p^2 \rangle = \langle (m x)^2 \rangle$ with $m$ the magnetization [note that this term comes from regarding the spin as a charge, and is not related to the physical magnetic dipole moment]. A magnon moves (to leading order) with a well-defined velocity $v$, so
\beq
\langle p^2 \rangle = v^2 t^2 \langle m^2 \rangle.
\eeq
On average, as we have noted, there is no magnetization so $\langle m \rangle = 0$. However, by central limiting arguments, the region through which the spin propagates on this timescale has ${\cal O}(\sqrt{vt})$ more $\uparrow$ domains than $\downarrow$ domains. Thus, the magnetization varies as $\langle m^2 \rangle \sim 1/\left| vt \right|$, giving us the result that
\beq
\langle p^2 \rangle \sim \left|v \right| t,
\eeq
which implies spin diffusion with a diffusion constant $\sim \left| v \right|$. 

\emph{Spin diffusion constant}.---Generalizing this away from the $\Delta \gg 1$ limit requires some nontrivial steps: the magnons are at high density, so cannot be treated as being dilute; also, magnon strings of all lengths are mobile, and must be accounted for. 
However, these issues can be addressed at the level of GHD~\cite{Doyon, Fagotti}: we assume that quasiparticles are in local equilibrium in an appropriate generalized Gibbs ensemble, and evaluate the dressed quasiparticle dispersion as well as the quasiparticle distribution function using data from the thermodynamic Bethe ansatz solution. Many of the results we will use here are presented in the Supplemental Material of Ref.~\cite{idmp}. We work at half-filling ($S_z=0$), and infinite temperature.
Under these assumptions, each string can be addressed separately, and the full dipole moment is a sum over $s$-strings, as follows~\cite{GHDII,PhysRevB.96.081118}
\beq\label{kin1}
\langle p^2 \rangle = t^2 \sum_s \int du \, v_s(u)^2 \rho_s(u) (1 - \theta_s(u))  (m_s(h,u))^2.
\eeq
Here, the discrete index $s$ runs over $s$-string quasiparticles, and $u$ denotes the rapidities of the quasiparticles; $\rho_s(u)$ is the density of that species of quasiparticle, $\theta_s(u)$ their occupation number, $v_s(u)$ their effective velocity~\cite{PhysRevLett.113.187203,Doyon, Fagotti}, and $m_s$ is the (dressed) magnetization of a quasiparticle in a state that is $h$ away from half-filling. Essentially all the terms entering~\eqref{kin1} are thermodynamic data, for which tractable Bethe ansatz results exist in the limit of infinite temperature. Note that we focused on the thermal fluctuations of the dressed magnetization and ignored the subleading fluctuations of the effective quasiparticle velocity~\cite{ghkv} since the velocity fluctuations multiply the average dressed magnetization, which vanishes at half-filling. Thus the leading fluctuation effect comes from fluctuations of the local background magnetization through which the quasiparticle is propagating. We expect $\langle p^2 \rangle $ to be related to the spin diffusion constant as $\frac{\langle p^2 \rangle}{\langle m^2 \rangle}= 2 D t$ with $\langle m^2 \rangle= \frac{1}{4}$ the spin susceptibility at infinite temperature.

We now proceed to simplify the expectation value in the infinite-temperature limit, using a Bethe ansatz result~\cite{idmp}
\beq\label{heq}
 m_s(h)^2  = \frac{1}{9} (s + 1)^4 \langle h^2 \rangle = \frac{1}{9} \frac{(s+1)^4}{|v_s(u)t|},
\eeq
with $h$ an effective small magnetic field felt by the quasiparticle over its finite trajectory $|v_s(u) t|$.
A key step here is that we choose to measure magnetization fluctuations over a distance $|v_s(u) t|$; once this is chosen, the quantitative value of the variance follows from being at infinite temperature. 

The justification for averaging over a distance $|v_s(u) t|$ is ultimately physical: the particle moves ballistically whereas (per previous rigorous results) spin is transported sub-ballistically; therefore, the motion of the background spin configuration is parametrically slower than the motion of the quasiparticle, and can thus be treated as frozen. Consequently the quasiparticle only experiences background fluctuations out to the distance to which it has moved. This feature is an important distinction between the present problem and that of quasiparticle front diffusion: there both the quasiparticle and its environment were moving ballistically, so one instead had to work with \emph{velocity differences} rather than just velocities~\cite{dbd1, ghkv}. It is also a crucial distinction between our approach and that of Ref.~\cite{idmp}, where the aim was to find a lower bound and therefore the averaging was always done over the largest meaningful distance scale in the problem, namely the Lieb-Robinson velocity~\cite{Lieb2004}. 

From this step onwards, the argument is straightforward. We use that $\rho_s(u) = \theta_s \rho^{\rm tot}_s(u) $, where $\rho^{\rm tot}_s(u)$ is a density of states factor, to rewrite Eq.~\eqref{kin1} as 
\beq\label{kin2}
\langle p^2 \rangle = t \sum_s \int du \, \rho^{\rm tot}_s(u) |v_s(u)| \theta_s (1 - \theta_s) \frac{(s + 1)^4}{9}.
\eeq
This yields a diffusion constant given by
\beq\label{kinD}
D= \frac{2}{9} \sum_s \theta_s (1 - \theta_s) (s + 1)^4 \int du \, \rho^{\rm tot}_s(u) |v_s(u)|,
\eeq
where we have used the fact that $\theta_s=\frac{1}{(1+s)^2}$ is independent of the rapidity $u$ at infinite temperature~\cite{Takahashi}. 

\emph{Superdiffusion for $\Delta=1$}.---The nature of the solution depends crucially on the $s$-dependence of $v_s(u)$. For $\Delta > 1$, large-$s$ strings have exponentially suppressed velocities and thus do not contribute to the dipole moment; the integral and the sum over $s$ converge and give a well-defined diffusion constant. For $\Delta = 1$, however, the maximum velocities fall off slowly with $s$ as $1/s$, leading to an algebraic decay of the integral
\beq
\int du \rho^{\rm tot}_s(u)|v_s(u)| \sim \frac{1}{s^2}.
\eeq
Meanwhile, for large $s$, $\theta_s \simeq 1/s^2$. Plugging in the expressions for these quantities we end up with a summand that has no $s$-dependence to leading order, and thus to an apparently divergent diffusion constant that scales as
$D \sim s^*$,
with $s^*$ a cutoff on the number of strings ($s< s^*$).
Anomalous diffusion is a consequence of the way this divergence gets cut off. 
Our argument that led to Eq.~\eqref{kin2} assumed that the ballistic motion of a string was faster than the rearrangement of its surroundings. 
However, this assumption must clearly fail at a given time $t$ for sufficiently heavy strings. We proceed as follows. First, we will assume that spin diffuses anomalously, with an anomalous exponent $\alpha$, i.e., that $D(t) \sim t^{\alpha}$. We will then compute $\alpha$ self-consistently. 

Assuming an exponent $\alpha$, we find that our assumption of primarily ballistic motion breaks down for strings such that for which $v_s t \simeq t/s \leq \sqrt{D t} = t^{(1 + \alpha)/2}$, the anomalous diffusion distance. 
We must therefore separate our sum at $s^* \simeq t^{(1-\alpha)/2}$ into two parts. The first part consists of light strings up to $s^*$; the second consists of yet heavier strings. The first set of terms, plugged into $D \sim t^\alpha \sim s^*$, immediately gives
\beq
\alpha = (1 - \alpha)/2 \Rightarrow \alpha = 1/3.
\eeq
This exponent can also be recovered by recognizing that the expression~\eqref{heq} is only valid for $s < h^{-1} \sim\sqrt{v_s t}$. Using the characteristic scaling $v_s \sim 1/s$, this yields $s < s^* \sim t^{1/3}$ in agreement with the previous argument.  

To complete this treatment we need to account for the heavy strings, $s > s^*$. 
These strings move ballistically but slowly, so that $v_s t \alt \sqrt{D(t) t}$.
Thus the net magnetization they experience on timescale $t$ is set, not by their ballistic motion, but by the anomalous diffusion length-scale of the background spin environment, $x(t) = \sqrt{D(t) t} \sim t^{2/3}$. Therefore, the characteristic net magnetic field heavy strings feel is $s$-independent, and is given by $h^{\mathrm{heavy}}(t) \sim 1/\sqrt{x(t)} \sim 1/s^* > 1/s$. 
Thus, strings with $s > s^*$ have a dressed magnetization that is effectively their bare magnetization $s$,
%
and each heavy string individually gives a ballistic contribution to $\langle p^2 \rangle$. The integral over the rapidity now scales as $\int du \rho^{\rm tot}_s(u)v_s(u)^2 \sim \frac{1}{s^3}$~\cite{idmp}, so the sum over strings larger than $s^*$ goes as $\sum \frac{1}{s^3}$ and is convergent at large $s$. This gives a contribution $t^2 \sum_{s>s^*} \frac{1}{s^3} \sim t^{4/3}$ to $\langle p^2 \rangle$. The contributions from light strings $s < s^*$ and $s > s^*$ are of the same magnitude. Finally, strings with $s \simeq s^*$ move ballistically over a distance $\sim t^{2/3}$. The contribution of each such string to $\langle p^2 \rangle$ scales as $t^{2-2/3}=t^{4/3}$, consistent with the two regimes $s < s^*$ and $s > s^*$.

We therefore conclude that the diffusion constant scales as
\beq
D(t) \sim  t^{1/3}.
\eeq
This expression is our main result. It predicts that spin is transported over a distance $t^{2/3} $ in a time $t$. This power-law behavior has been observed in numerical studies~\cite{lzp}, and has remained a mystery until now.

\emph{Easy-axis phase $\Delta > 1$}.---We now turn to the implications of our argument for transport in the easy-axis phase of the XXZ chain away from the isotropic point. First, we discuss the vicinity of the isotropic point, $\Delta \rightarrow 1^+$. Here, the sum over strings is cut off exponentially at the scale $s^* \sim 1/\eta$, where $\eta = \cosh^{-1}(\Delta)$. Since the diffusion constant goes as $s^*$, it follows that it diverges as $D  \sim 1/\sqrt{\Delta - 1}$ near the isotropic limit.  For $\Delta > 1$, our analysis would predict a crossover from anomalous to normal diffusion on a timescale that scales as $t^* \sim (\Delta - 1)^{-3/2}$. 

Our expressions Eq.~\eqref{kin2} and Eq.~\eqref{kinD} were derived under mild assumptions that are valid throughout the easy-axis regime. Thus, they can be used to derive analytic expressions for the infinite-temperature diffusion constant for $\Delta > 1$. Extracting an analytic expression for the spin diffusion constant has remained a long-standing open problem in the field, which the kinetic approach allows us to address.

The integral in Eq.~\eqref{kinD} can be computed analytically for any $\Delta=\cosh \eta >1$. This allows us to find a closed-form expression for the diffusion constant
\beq
\label{eqDiffusionConstant}
D = \frac{2 \sinh \eta}{9 \pi} \sum_{s=1}^\infty (1+s) \left[\frac{s+2}{\sinh \eta s} - \frac{s}{\sinh \eta (s+2)} \right].
\eeq
The diffusion constant as a function of $\Delta$ is plotted in Fig.~\ref{FigD}. At large anisotropy $\Delta \gg 1$ one can evaluate the sum explicitly, as only the $s = 1$ string contributes in that limit. We find

\beq
\lim_{\Delta \to \infty}D = \frac{4}{3 \pi} \simeq 0.4244,
\eeq
consistent with numerical results~\cite{PhysRevB.89.075139,1367-2630-19-3-033027} and with the  bound found in Ref.~\cite{idmp}. 

\emph{Discussion}.---Our results were derived specifically for spin diffusion in an infinite-temperature state, but our prediction of superdiffusion with $D(t) \sim t^{1/3}$ for the isotropic Heinsenberg chain can be readily extended to any non-zero temperature $T>0$ using the results of Ref.~\cite{idmp}. Our kinetic approach to compute the diffusion constant in the easy-axis regime can also be generalized to any temperature; in particular, it would be interesting to recover the predictions of Refs.~\cite{PhysRevLett.78.943,PhysRevB.57.8307,PhysRevLett.95.187201} at low temperature. 

Our predictions for (super)diffusion are directly testable in ultracold atomic experiments~\cite{fukuhara2013}. However, many numerical studies of the XXZ model work with a different setup, in which the left and right halves of a system are initialized at different values of magnetization, and the spin current through the middle of the system is measured~\cite{lzp}. For thermal systems these procedures are equivalent; however, showing that this remains the case in the present setting requires further justification. When the bias $\delta \mu$ is sufficiently small~\cite{lzp}, this is straightforward. A quasiparticle traveling for a time $t$ picks up an average magnetization $\sim t \delta \mu $, with a sign that depends on which half of the system it is in, in addition to its fluctuations. Thus the magnetization picks up a drift in addition to its (normal or anomalous) diffusion, just as a conventional diffusive particle would. The case of larger $\delta \mu$ can also be addressed using the present framework, but requires a detailed analysis of the $\mu$-dependence of the dressed magnetization; this will be addressed elsewhere. Our argument requires non-vanishing thermal fluctuations, and does not apply to zero-entropy initial states~\cite{1742-5468-2017-10-103108,PhysRevB.96.195151,PhysRevB.97.081111,2018arXiv181008227B}.

Note that the mechanism for superdiffusion described above in integrable systems is quite different from that in random classical systems. The fact that the spin front observed numerically is a rescaled error function~\cite{lzp} and not, e.g., a Levy flight, might suggest that the dynamics of the dipole moment on timescale $t$ is dominated by the random walk associated with the strings that have the largest dipole moments on that timescale. The kinetic approach here can be extended to compute higher moments $\langle p^n \rangle$, and thus the full shape of the magnetization front; however, unlike $\langle p^2 \rangle$, higher moments will also involve fluctuations of the quasiparticle velocities. We defer a detailed treatment of this question to future work.


While we focused on the overall spread of the initial spin packet, an interesting question for future work is the nature of the spin structure factor $C(x,t) \equiv \langle \sigma^z(x,t) \sigma^z(0,0) \rangle$ more generally. Even though contributions from large, heavy strings do not affect the overall size of the wavepacket, they might affect the shape of $C(x,t)$ near the origin, and can potentially dominate the local autocorrelation function, leading to anomalous behavior~\footnote{S. Gopalakrishnan and V. Alba, in preparation.}. The form of this structure factor, and thus of the frequency and wavevector dependent conductivity in the hydrodynamic limit, can in principle be understood using the kinetic approach outlined here, and is a promising subject for future study.

Finally, we remark that an implication of our theory is that operator spreading should presumably remain ballistic, with diffusive broadening~\cite{ghkv}, in the XXX model: the quasiparticle trajectories do not change their character at the isotropic point; instead, the unconventional physics is due to anomalous fluctuations of the magnetization along these trajectories. 

\emph{Acknowledgments}.---The authors thank Vir Bulchandani, Vadim Oganesyan, and Thomas Scaffidi for helpful discussions, and David Huse and Vedika Khemani for collaborations on related work. We also thank Vir Bulchandani, Jacopo De Nardis and Subir Sachdev for useful comments on the manuscript. This work was supported by NSF Grant No. DMR-1653271 (S.G.), and US Department of Energy, Office of Science, Basic Energy Sciences, under Award No. DE-SC0019168 (R.V.).

\emph{Note}.---While we were writing this manuscript, the paper~\cite{dbd2} appeared on Arxiv. While Ref.~\cite{dbd2} does not address superdiffusion, the authors use their previous approach~\cite{dbd1} to evaluate the spin diffusion constant in the easy-axis (``gapped'') phase. The numerical evaluation of the formula of Ref.~\cite{dbd2} are in perfect agreement with our analytic result~\eqref{eqDiffusionConstant}.

\bibliography{References}

\begin{thebibliography}{71}%
\makeatletter
\providecommand \@ifxundefined [1]{%
 \@ifx{#1\undefined}
}%
\providecommand \@ifnum [1]{%
 \ifnum #1\expandafter \@firstoftwo
 \else \expandafter \@secondoftwo
 \fi
}%
\providecommand \@ifx [1]{%
 \ifx #1\expandafter \@firstoftwo
 \else \expandafter \@secondoftwo
 \fi
}%
\providecommand \natexlab [1]{#1}%
\providecommand \enquote  [1]{``#1''}%
\providecommand \bibnamefont  [1]{#1}%
\providecommand \bibfnamefont [1]{#1}%
\providecommand \citenamefont [1]{#1}%
\providecommand \href@noop [0]{\@secondoftwo}%
\providecommand \href [0]{\begingroup \@sanitize@url \@href}%
\providecommand \@href[1]{\@@startlink{#1}\@@href}%
\providecommand \@@href[1]{\endgroup#1\@@endlink}%
\providecommand \@sanitize@url [0]{\catcode `\\12\catcode `\$12\catcode
  `\&12\catcode `\#12\catcode `\^12\catcode `\_12\catcode `\%12\relax}%
\providecommand \@@startlink[1]{}%
\providecommand \@@endlink[0]{}%
\providecommand \url  [0]{\begingroup\@sanitize@url \@url }%
\providecommand \@url [1]{\endgroup\@href {#1}{\urlprefix }}%
\providecommand \urlprefix  [0]{URL }%
\providecommand \Eprint [0]{\href }%
\providecommand \doibase [0]{http://dx.doi.org/}%
\providecommand \selectlanguage [0]{\@gobble}%
\providecommand \bibinfo  [0]{\@secondoftwo}%
\providecommand \bibfield  [0]{\@secondoftwo}%
\providecommand \translation [1]{[#1]}%
\providecommand \BibitemOpen [0]{}%
\providecommand \bibitemStop [0]{}%
\providecommand \bibitemNoStop [0]{.\EOS\space}%
\providecommand \EOS [0]{\spacefactor3000\relax}%
\providecommand \BibitemShut  [1]{\csname bibitem#1\endcsname}%
\let\auto@bib@innerbib\@empty
\bibitem [{\citenamefont {Rigol}\ \emph {et~al.}(2007)\citenamefont {Rigol},
  \citenamefont {Dunjko}, \citenamefont {Yurovsky},\ and\ \citenamefont
  {Olshanii}}]{PhysRevLett.98.050405}%
  \BibitemOpen
  \bibfield  {author} {\bibinfo {author} {\bibfnamefont {M.}~\bibnamefont
  {Rigol}}, \bibinfo {author} {\bibfnamefont {V.}~\bibnamefont {Dunjko}},
  \bibinfo {author} {\bibfnamefont {V.}~\bibnamefont {Yurovsky}}, \ and\
  \bibinfo {author} {\bibfnamefont {M.}~\bibnamefont {Olshanii}},\ }\href
  {\doibase 10.1103/PhysRevLett.98.050405} {\bibfield  {journal} {\bibinfo
  {journal} {Phys. Rev. Lett.}\ }\textbf {\bibinfo {volume} {98}},\ \bibinfo
  {pages} {050405} (\bibinfo {year} {2007})}\BibitemShut {NoStop}%
\bibitem [{\citenamefont {Rigol}\ \emph {et~al.}(2008)\citenamefont {Rigol},
  \citenamefont {Dunjko},\ and\ \citenamefont {Olshanii}}]{Rigol:2008kq}%
  \BibitemOpen
  \bibfield  {author} {\bibinfo {author} {\bibfnamefont {M.}~\bibnamefont
  {Rigol}}, \bibinfo {author} {\bibfnamefont {V.}~\bibnamefont {Dunjko}}, \
  and\ \bibinfo {author} {\bibfnamefont {M.}~\bibnamefont {Olshanii}},\ }\href
  {http://dx.doi.org/10.1038/nature06838} {\bibfield  {journal} {\bibinfo
  {journal} {Nature}\ }\textbf {\bibinfo {volume} {452}},\ \bibinfo {pages}
  {854} (\bibinfo {year} {2008})}\BibitemShut {NoStop}%
\bibitem [{\citenamefont {Calabrese}\ and\ \citenamefont
  {Cardy}(2006)}]{Calabrese:2006}%
  \BibitemOpen
  \bibfield  {author} {\bibinfo {author} {\bibfnamefont {P.}~\bibnamefont
  {Calabrese}}\ and\ \bibinfo {author} {\bibfnamefont {J.}~\bibnamefont
  {Cardy}},\ }\href@noop {} {\bibfield  {journal} {\bibinfo  {journal}
  {Physical Review Letters}\ }\textbf {\bibinfo {volume} {96}},\ \bibinfo
  {pages} {136801} (\bibinfo {year} {2006})}\BibitemShut {NoStop}%
\bibitem [{\citenamefont {Prosen}(2011)}]{PhysRevLett.106.217206}%
  \BibitemOpen
  \bibfield  {author} {\bibinfo {author} {\bibfnamefont {T.~c.~v.}\
  \bibnamefont {Prosen}},\ }\href {\doibase 10.1103/PhysRevLett.106.217206}
  {\bibfield  {journal} {\bibinfo  {journal} {Phys. Rev. Lett.}\ }\textbf
  {\bibinfo {volume} {106}},\ \bibinfo {pages} {217206} (\bibinfo {year}
  {2011})}\BibitemShut {NoStop}%
\bibitem [{\citenamefont {Caux}\ and\ \citenamefont
  {Essler}(2013)}]{PhysRevLett.110.257203}%
  \BibitemOpen
  \bibfield  {author} {\bibinfo {author} {\bibfnamefont {J.-S.}\ \bibnamefont
  {Caux}}\ and\ \bibinfo {author} {\bibfnamefont {F.~H.~L.}\ \bibnamefont
  {Essler}},\ }\href {\doibase 10.1103/PhysRevLett.110.257203} {\bibfield
  {journal} {\bibinfo  {journal} {Phys. Rev. Lett.}\ }\textbf {\bibinfo
  {volume} {110}},\ \bibinfo {pages} {257203} (\bibinfo {year}
  {2013})}\BibitemShut {NoStop}%
\bibitem [{\citenamefont {Wouters}\ \emph {et~al.}(2014)\citenamefont
  {Wouters}, \citenamefont {De~Nardis}, \citenamefont {Brockmann},
  \citenamefont {Fioretto}, \citenamefont {Rigol},\ and\ \citenamefont
  {Caux}}]{PhysRevLett.113.117202}%
  \BibitemOpen
  \bibfield  {author} {\bibinfo {author} {\bibfnamefont {B.}~\bibnamefont
  {Wouters}}, \bibinfo {author} {\bibfnamefont {J.}~\bibnamefont {De~Nardis}},
  \bibinfo {author} {\bibfnamefont {M.}~\bibnamefont {Brockmann}}, \bibinfo
  {author} {\bibfnamefont {D.}~\bibnamefont {Fioretto}}, \bibinfo {author}
  {\bibfnamefont {M.}~\bibnamefont {Rigol}}, \ and\ \bibinfo {author}
  {\bibfnamefont {J.-S.}\ \bibnamefont {Caux}},\ }\href {\doibase
  10.1103/PhysRevLett.113.117202} {\bibfield  {journal} {\bibinfo  {journal}
  {Phys. Rev. Lett.}\ }\textbf {\bibinfo {volume} {113}},\ \bibinfo {pages}
  {117202} (\bibinfo {year} {2014})}\BibitemShut {NoStop}%
\bibitem [{\citenamefont {Ilievski}\ \emph {et~al.}(2015)\citenamefont
  {Ilievski}, \citenamefont {De~Nardis}, \citenamefont {Wouters}, \citenamefont
  {Caux}, \citenamefont {Essler},\ and\ \citenamefont
  {Prosen}}]{PhysRevLett.115.157201}%
  \BibitemOpen
  \bibfield  {author} {\bibinfo {author} {\bibfnamefont {E.}~\bibnamefont
  {Ilievski}}, \bibinfo {author} {\bibfnamefont {J.}~\bibnamefont {De~Nardis}},
  \bibinfo {author} {\bibfnamefont {B.}~\bibnamefont {Wouters}}, \bibinfo
  {author} {\bibfnamefont {J.-S.}\ \bibnamefont {Caux}}, \bibinfo {author}
  {\bibfnamefont {F.~H.~L.}\ \bibnamefont {Essler}}, \ and\ \bibinfo {author}
  {\bibfnamefont {T.}~\bibnamefont {Prosen}},\ }\href {\doibase
  10.1103/PhysRevLett.115.157201} {\bibfield  {journal} {\bibinfo  {journal}
  {Phys. Rev. Lett.}\ }\textbf {\bibinfo {volume} {115}},\ \bibinfo {pages}
  {157201} (\bibinfo {year} {2015})}\BibitemShut {NoStop}%
\bibitem [{\citenamefont {Ilievski}\ \emph {et~al.}(2016)\citenamefont
  {Ilievski}, \citenamefont {Medenjak}, \citenamefont {Prosen},\ and\
  \citenamefont {Zadnik}}]{2016arXiv160300440I}%
  \BibitemOpen
  \bibfield  {author} {\bibinfo {author} {\bibfnamefont {E.}~\bibnamefont
  {Ilievski}}, \bibinfo {author} {\bibfnamefont {M.}~\bibnamefont {Medenjak}},
  \bibinfo {author} {\bibfnamefont {T.}~\bibnamefont {Prosen}}, \ and\ \bibinfo
  {author} {\bibfnamefont {L.}~\bibnamefont {Zadnik}},\ }\href
  {http://stacks.iop.org/1742-5468/2016/i=6/a=064008} {\bibfield  {journal}
  {\bibinfo  {journal} {Journal of Statistical Mechanics: Theory and
  Experiment}\ }\textbf {\bibinfo {volume} {2016}},\ \bibinfo {pages} {064008}
  (\bibinfo {year} {2016})}\BibitemShut {NoStop}%
\bibitem [{\citenamefont {Calabrese}\ \emph {et~al.}(2016)\citenamefont
  {Calabrese}, \citenamefont {Essler},\ and\ \citenamefont
  {Mussardo}}]{1742-5468-2016-6-064001}%
  \BibitemOpen
  \bibfield  {author} {\bibinfo {author} {\bibfnamefont {P.}~\bibnamefont
  {Calabrese}}, \bibinfo {author} {\bibfnamefont {F.~H.~L.}\ \bibnamefont
  {Essler}}, \ and\ \bibinfo {author} {\bibfnamefont {G.}~\bibnamefont
  {Mussardo}},\ }\href {http://stacks.iop.org/1742-5468/2016/i=6/a=064001}
  {\bibfield  {journal} {\bibinfo  {journal} {Journal of Statistical Mechanics:
  Theory and Experiment}\ }\textbf {\bibinfo {volume} {2016}},\ \bibinfo
  {pages} {064001} (\bibinfo {year} {2016})}\BibitemShut {NoStop}%
\bibitem [{\citenamefont {Essler}\ and\ \citenamefont
  {Fagotti}(2016)}]{1742-5468-2016-6-064002}%
  \BibitemOpen
  \bibfield  {author} {\bibinfo {author} {\bibfnamefont {F.~H.~L.}\
  \bibnamefont {Essler}}\ and\ \bibinfo {author} {\bibfnamefont
  {M.}~\bibnamefont {Fagotti}},\ }\href
  {http://stacks.iop.org/1742-5468/2016/i=6/a=064002} {\bibfield  {journal}
  {\bibinfo  {journal} {Journal of Statistical Mechanics: Theory and
  Experiment}\ }\textbf {\bibinfo {volume} {2016}},\ \bibinfo {pages} {064002}
  (\bibinfo {year} {2016})}\BibitemShut {NoStop}%
\bibitem [{\citenamefont {Vasseur}\ and\ \citenamefont
  {Moore}(2016)}]{1742-5468-2016-6-064010}%
  \BibitemOpen
  \bibfield  {author} {\bibinfo {author} {\bibfnamefont {R.}~\bibnamefont
  {Vasseur}}\ and\ \bibinfo {author} {\bibfnamefont {J.~E.}\ \bibnamefont
  {Moore}},\ }\href {http://stacks.iop.org/1742-5468/2016/i=6/a=064010}
  {\bibfield  {journal} {\bibinfo  {journal} {Journal of Statistical Mechanics:
  Theory and Experiment}\ }\textbf {\bibinfo {volume} {2016}},\ \bibinfo
  {pages} {064010} (\bibinfo {year} {2016})}\BibitemShut {NoStop}%
\bibitem [{\citenamefont {Vidmar}\ and\ \citenamefont
  {Rigol}(2016)}]{1742-5468-2016-6-064007}%
  \BibitemOpen
  \bibfield  {author} {\bibinfo {author} {\bibfnamefont {L.}~\bibnamefont
  {Vidmar}}\ and\ \bibinfo {author} {\bibfnamefont {M.}~\bibnamefont {Rigol}},\
  }\href {http://stacks.iop.org/1742-5468/2016/i=6/a=064007} {\bibfield
  {journal} {\bibinfo  {journal} {Journal of Statistical Mechanics: Theory and
  Experiment}\ }\textbf {\bibinfo {volume} {2016}},\ \bibinfo {pages} {064007}
  (\bibinfo {year} {2016})}\BibitemShut {NoStop}%
\bibitem [{\citenamefont {Fagotti}\ \emph {et~al.}(2014)\citenamefont
  {Fagotti}, \citenamefont {Collura}, \citenamefont {Essler},\ and\
  \citenamefont {Calabrese}}]{PhysRevB.89.125101}%
  \BibitemOpen
  \bibfield  {author} {\bibinfo {author} {\bibfnamefont {M.}~\bibnamefont
  {Fagotti}}, \bibinfo {author} {\bibfnamefont {M.}~\bibnamefont {Collura}},
  \bibinfo {author} {\bibfnamefont {F.~H.~L.}\ \bibnamefont {Essler}}, \ and\
  \bibinfo {author} {\bibfnamefont {P.}~\bibnamefont {Calabrese}},\ }\href
  {\doibase 10.1103/PhysRevB.89.125101} {\bibfield  {journal} {\bibinfo
  {journal} {Phys. Rev. B}\ }\textbf {\bibinfo {volume} {89}},\ \bibinfo
  {pages} {125101} (\bibinfo {year} {2014})}\BibitemShut {NoStop}%
\bibitem [{\citenamefont {Alba}\ and\ \citenamefont
  {Calabrese}(2017)}]{alba2017entanglement}%
  \BibitemOpen
  \bibfield  {author} {\bibinfo {author} {\bibfnamefont {V.}~\bibnamefont
  {Alba}}\ and\ \bibinfo {author} {\bibfnamefont {P.}~\bibnamefont
  {Calabrese}},\ }\href@noop {} {\bibfield  {journal} {\bibinfo  {journal}
  {Proceedings of the National Academy of Sciences}\ }\textbf {\bibinfo
  {volume} {114}},\ \bibinfo {pages} {7947} (\bibinfo {year}
  {2017})}\BibitemShut {NoStop}%
\bibitem [{\citenamefont {Kinoshita}\ \emph {et~al.}(2006)\citenamefont
  {Kinoshita}, \citenamefont {Wenger},\ and\ \citenamefont
  {Weiss}}]{kinoshita}%
  \BibitemOpen
  \bibfield  {author} {\bibinfo {author} {\bibfnamefont {T.}~\bibnamefont
  {Kinoshita}}, \bibinfo {author} {\bibfnamefont {T.}~\bibnamefont {Wenger}}, \
  and\ \bibinfo {author} {\bibfnamefont {D.}~\bibnamefont {Weiss}},\
  }\href@noop {} {\bibfield  {journal} {\bibinfo  {journal} {Nature}\ }\textbf
  {\bibinfo {volume} {440}},\ \bibinfo {pages} {900} (\bibinfo {year}
  {2006})}\BibitemShut {NoStop}%
\bibitem [{\citenamefont {Gring}\ \emph {et~al.}(2012)\citenamefont {Gring},
  \citenamefont {Kuhnert}, \citenamefont {Langen}, \citenamefont {Kitagawa},
  \citenamefont {Rauer}, \citenamefont {Schreitl}, \citenamefont {Mazets},
  \citenamefont {Smith}, \citenamefont {Demler},\ and\ \citenamefont
  {Schmiedmayer}}]{gring}%
  \BibitemOpen
  \bibfield  {author} {\bibinfo {author} {\bibfnamefont {M.}~\bibnamefont
  {Gring}}, \bibinfo {author} {\bibfnamefont {M.}~\bibnamefont {Kuhnert}},
  \bibinfo {author} {\bibfnamefont {T.}~\bibnamefont {Langen}}, \bibinfo
  {author} {\bibfnamefont {T.}~\bibnamefont {Kitagawa}}, \bibinfo {author}
  {\bibfnamefont {B.}~\bibnamefont {Rauer}}, \bibinfo {author} {\bibfnamefont
  {M.}~\bibnamefont {Schreitl}}, \bibinfo {author} {\bibfnamefont
  {I.}~\bibnamefont {Mazets}}, \bibinfo {author} {\bibfnamefont {D.~A.}\
  \bibnamefont {Smith}}, \bibinfo {author} {\bibfnamefont {E.}~\bibnamefont
  {Demler}}, \ and\ \bibinfo {author} {\bibfnamefont {J.}~\bibnamefont
  {Schmiedmayer}},\ }\href@noop {} {\bibfield  {journal} {\bibinfo  {journal}
  {Science}\ ,\ \bibinfo {pages} {1224953}} (\bibinfo {year}
  {2012})}\BibitemShut {NoStop}%
\bibitem [{\citenamefont {Tang}\ \emph {et~al.}(2018)\citenamefont {Tang},
  \citenamefont {Kao}, \citenamefont {Li}, \citenamefont {Seo}, \citenamefont
  {Mallayya}, \citenamefont {Rigol}, \citenamefont {Gopalakrishnan},\ and\
  \citenamefont {Lev}}]{tang2018}%
  \BibitemOpen
  \bibfield  {author} {\bibinfo {author} {\bibfnamefont {Y.}~\bibnamefont
  {Tang}}, \bibinfo {author} {\bibfnamefont {W.}~\bibnamefont {Kao}}, \bibinfo
  {author} {\bibfnamefont {K.-Y.}\ \bibnamefont {Li}}, \bibinfo {author}
  {\bibfnamefont {S.}~\bibnamefont {Seo}}, \bibinfo {author} {\bibfnamefont
  {K.}~\bibnamefont {Mallayya}}, \bibinfo {author} {\bibfnamefont
  {M.}~\bibnamefont {Rigol}}, \bibinfo {author} {\bibfnamefont
  {S.}~\bibnamefont {Gopalakrishnan}}, \ and\ \bibinfo {author} {\bibfnamefont
  {B.~L.}\ \bibnamefont {Lev}},\ }\href {\doibase 10.1103/PhysRevX.8.021030}
  {\bibfield  {journal} {\bibinfo  {journal} {Phys. Rev. X}\ }\textbf {\bibinfo
  {volume} {8}},\ \bibinfo {pages} {021030} (\bibinfo {year}
  {2018})}\BibitemShut {NoStop}%
\bibitem [{\citenamefont {Erne}\ \emph {et~al.}(2018)\citenamefont {Erne},
  \citenamefont {B{\"u}cker}, \citenamefont {Gasenzer}, \citenamefont
  {Berges},\ and\ \citenamefont {Schmiedmayer}}]{erne2018}%
  \BibitemOpen
  \bibfield  {author} {\bibinfo {author} {\bibfnamefont {S.}~\bibnamefont
  {Erne}}, \bibinfo {author} {\bibfnamefont {R.}~\bibnamefont {B{\"u}cker}},
  \bibinfo {author} {\bibfnamefont {T.}~\bibnamefont {Gasenzer}}, \bibinfo
  {author} {\bibfnamefont {J.}~\bibnamefont {Berges}}, \ and\ \bibinfo {author}
  {\bibfnamefont {J.}~\bibnamefont {Schmiedmayer}},\ }\href@noop {} {\bibfield
  {journal} {\bibinfo  {journal} {Nature}\ }\textbf {\bibinfo {volume} {563}},\
  \bibinfo {pages} {225} (\bibinfo {year} {2018})}\BibitemShut {NoStop}%
\bibitem [{\citenamefont {Zundel}\ \emph {et~al.}(2018)\citenamefont {Zundel},
  \citenamefont {Wilson}, \citenamefont {Malvania}, \citenamefont {Xia},
  \citenamefont {Riou},\ and\ \citenamefont {Weiss}}]{zundel2018}%
  \BibitemOpen
  \bibfield  {author} {\bibinfo {author} {\bibfnamefont {L.~A.}\ \bibnamefont
  {Zundel}}, \bibinfo {author} {\bibfnamefont {J.~M.}\ \bibnamefont {Wilson}},
  \bibinfo {author} {\bibfnamefont {N.}~\bibnamefont {Malvania}}, \bibinfo
  {author} {\bibfnamefont {L.}~\bibnamefont {Xia}}, \bibinfo {author}
  {\bibfnamefont {J.-F.}\ \bibnamefont {Riou}}, \ and\ \bibinfo {author}
  {\bibfnamefont {D.~S.}\ \bibnamefont {Weiss}},\ }\href@noop {} {\bibfield
  {journal} {\bibinfo  {journal} {arXiv preprint arXiv:1810.00120}\ } (\bibinfo
  {year} {2018})}\BibitemShut {NoStop}%
\bibitem [{\citenamefont {Langer}\ \emph {et~al.}(2009)\citenamefont {Langer},
  \citenamefont {Heidrich-Meisner}, \citenamefont {Gemmer}, \citenamefont
  {McCulloch},\ and\ \citenamefont {Schollw\"ock}}]{PhysRevB.79.214409}%
  \BibitemOpen
  \bibfield  {author} {\bibinfo {author} {\bibfnamefont {S.}~\bibnamefont
  {Langer}}, \bibinfo {author} {\bibfnamefont {F.}~\bibnamefont
  {Heidrich-Meisner}}, \bibinfo {author} {\bibfnamefont {J.}~\bibnamefont
  {Gemmer}}, \bibinfo {author} {\bibfnamefont {I.~P.}\ \bibnamefont
  {McCulloch}}, \ and\ \bibinfo {author} {\bibfnamefont {U.}~\bibnamefont
  {Schollw\"ock}},\ }\href {\doibase 10.1103/PhysRevB.79.214409} {\bibfield
  {journal} {\bibinfo  {journal} {Phys. Rev. B}\ }\textbf {\bibinfo {volume}
  {79}},\ \bibinfo {pages} {214409} (\bibinfo {year} {2009})}\BibitemShut
  {NoStop}%
\bibitem [{\citenamefont {Steinigeweg}\ and\ \citenamefont
  {Gemmer}(2009)}]{PhysRevB.80.184402}%
  \BibitemOpen
  \bibfield  {author} {\bibinfo {author} {\bibfnamefont {R.}~\bibnamefont
  {Steinigeweg}}\ and\ \bibinfo {author} {\bibfnamefont {J.}~\bibnamefont
  {Gemmer}},\ }\href {\doibase 10.1103/PhysRevB.80.184402} {\bibfield
  {journal} {\bibinfo  {journal} {Phys. Rev. B}\ }\textbf {\bibinfo {volume}
  {80}},\ \bibinfo {pages} {184402} (\bibinfo {year} {2009})}\BibitemShut
  {NoStop}%
\bibitem [{\citenamefont {Prosen}\ and\ \citenamefont {{\v Z}nidari{\v
  c}}(2009)}]{1742-5468-2009-02-P02035}%
  \BibitemOpen
  \bibfield  {author} {\bibinfo {author} {\bibfnamefont {T.}~\bibnamefont
  {Prosen}}\ and\ \bibinfo {author} {\bibfnamefont {M.}~\bibnamefont {{\v
  Z}nidari{\v c}}},\ }\href
  {http://stacks.iop.org/1742-5468/2009/i=02/a=P02035} {\bibfield  {journal}
  {\bibinfo  {journal} {Journal of Statistical Mechanics: Theory and
  Experiment}\ }\textbf {\bibinfo {volume} {2009}},\ \bibinfo {pages} {P02035}
  (\bibinfo {year} {2009})}\BibitemShut {NoStop}%
\bibitem [{\citenamefont {\ifmmode \check{Z}\else
  \v{Z}\fi{}nidari\ifmmode~\check{c}\else
  \v{c}\fi{}}(2011)}]{PhysRevLett.106.220601}%
  \BibitemOpen
  \bibfield  {author} {\bibinfo {author} {\bibfnamefont {M.}~\bibnamefont
  {\ifmmode \check{Z}\else \v{Z}\fi{}nidari\ifmmode~\check{c}\else
  \v{c}\fi{}}},\ }\href {\doibase 10.1103/PhysRevLett.106.220601} {\bibfield
  {journal} {\bibinfo  {journal} {Phys. Rev. Lett.}\ }\textbf {\bibinfo
  {volume} {106}},\ \bibinfo {pages} {220601} (\bibinfo {year}
  {2011})}\BibitemShut {NoStop}%
\bibitem [{\citenamefont {Steinigeweg}\ and\ \citenamefont
  {Brenig}(2011)}]{PhysRevLett.107.250602}%
  \BibitemOpen
  \bibfield  {author} {\bibinfo {author} {\bibfnamefont {R.}~\bibnamefont
  {Steinigeweg}}\ and\ \bibinfo {author} {\bibfnamefont {W.}~\bibnamefont
  {Brenig}},\ }\href {\doibase 10.1103/PhysRevLett.107.250602} {\bibfield
  {journal} {\bibinfo  {journal} {Phys. Rev. Lett.}\ }\textbf {\bibinfo
  {volume} {107}},\ \bibinfo {pages} {250602} (\bibinfo {year}
  {2011})}\BibitemShut {NoStop}%
\bibitem [{\citenamefont {Ljubotina}\ \emph {et~al.}(2017)\citenamefont
  {Ljubotina}, \citenamefont {{\v Z}nidari{\v c}},\ and\ \citenamefont
  {Prosen}}]{lzp}%
  \BibitemOpen
  \bibfield  {author} {\bibinfo {author} {\bibfnamefont {M.}~\bibnamefont
  {Ljubotina}}, \bibinfo {author} {\bibfnamefont {M.}~\bibnamefont {{\v
  Z}nidari{\v c}}}, \ and\ \bibinfo {author} {\bibfnamefont {T.}~\bibnamefont
  {Prosen}},\ }\href {http://dx.doi.org/10.1038/ncomms16117} {\bibfield
  {journal} {\bibinfo  {journal} {Nature Communications}\ }\textbf {\bibinfo
  {volume} {8}},\ \bibinfo {pages} {16117 EP } (\bibinfo {year}
  {2017})}\BibitemShut {NoStop}%
\bibitem [{\citenamefont {{Ilievski}}\ \emph {et~al.}(2018)\citenamefont
  {{Ilievski}}, \citenamefont {{De Nardis}}, \citenamefont {{Medenjak}},\ and\
  \citenamefont {{Prosen}}}]{idmp}%
  \BibitemOpen
  \bibfield  {author} {\bibinfo {author} {\bibfnamefont {E.}~\bibnamefont
  {{Ilievski}}}, \bibinfo {author} {\bibfnamefont {J.}~\bibnamefont {{De
  Nardis}}}, \bibinfo {author} {\bibfnamefont {M.}~\bibnamefont {{Medenjak}}},
  \ and\ \bibinfo {author} {\bibfnamefont {T.}~\bibnamefont {{Prosen}}},\
  }\href@noop {} {\bibfield  {journal} {\bibinfo  {journal} {ArXiv e-prints}\ }
  (\bibinfo {year} {2018})},\ \Eprint {http://arxiv.org/abs/1806.03288}
  {arXiv:1806.03288 [cond-mat.stat-mech]} \BibitemShut {NoStop}%
\bibitem [{\citenamefont {S{\'a}nchez}\ \emph {et~al.}(2018)\citenamefont
  {S{\'a}nchez}, \citenamefont {Varma},\ and\ \citenamefont
  {Oganesyan}}]{sanchez2018anomalous}%
  \BibitemOpen
  \bibfield  {author} {\bibinfo {author} {\bibfnamefont {R.~J.}\ \bibnamefont
  {S{\'a}nchez}}, \bibinfo {author} {\bibfnamefont {V.~K.}\ \bibnamefont
  {Varma}}, \ and\ \bibinfo {author} {\bibfnamefont {V.}~\bibnamefont
  {Oganesyan}},\ }\href@noop {} {\bibfield  {journal} {\bibinfo  {journal}
  {Physical Review B}\ }\textbf {\bibinfo {volume} {98}},\ \bibinfo {pages}
  {054415} (\bibinfo {year} {2018})}\BibitemShut {NoStop}%
\bibitem [{\citenamefont {Castro-Alvaredo}\ \emph {et~al.}(2016)\citenamefont
  {Castro-Alvaredo}, \citenamefont {Doyon},\ and\ \citenamefont
  {Yoshimura}}]{Doyon}%
  \BibitemOpen
  \bibfield  {author} {\bibinfo {author} {\bibfnamefont {O.~A.}\ \bibnamefont
  {Castro-Alvaredo}}, \bibinfo {author} {\bibfnamefont {B.}~\bibnamefont
  {Doyon}}, \ and\ \bibinfo {author} {\bibfnamefont {T.}~\bibnamefont
  {Yoshimura}},\ }\href {\doibase 10.1103/PhysRevX.6.041065} {\bibfield
  {journal} {\bibinfo  {journal} {Phys. Rev. X}\ }\textbf {\bibinfo {volume}
  {6}},\ \bibinfo {pages} {041065} (\bibinfo {year} {2016})}\BibitemShut
  {NoStop}%
\bibitem [{\citenamefont {Bertini}\ \emph {et~al.}(2016)\citenamefont
  {Bertini}, \citenamefont {Collura}, \citenamefont {De~Nardis},\ and\
  \citenamefont {Fagotti}}]{Fagotti}%
  \BibitemOpen
  \bibfield  {author} {\bibinfo {author} {\bibfnamefont {B.}~\bibnamefont
  {Bertini}}, \bibinfo {author} {\bibfnamefont {M.}~\bibnamefont {Collura}},
  \bibinfo {author} {\bibfnamefont {J.}~\bibnamefont {De~Nardis}}, \ and\
  \bibinfo {author} {\bibfnamefont {M.}~\bibnamefont {Fagotti}},\ }\href
  {\doibase 10.1103/PhysRevLett.117.207201} {\bibfield  {journal} {\bibinfo
  {journal} {Phys. Rev. Lett.}\ }\textbf {\bibinfo {volume} {117}},\ \bibinfo
  {pages} {207201} (\bibinfo {year} {2016})}\BibitemShut {NoStop}%
\bibitem [{\citenamefont {Doyon}\ and\ \citenamefont
  {Yoshimura}(2017)}]{SciPostPhys.2.2.014}%
  \BibitemOpen
  \bibfield  {author} {\bibinfo {author} {\bibfnamefont {B.}~\bibnamefont
  {Doyon}}\ and\ \bibinfo {author} {\bibfnamefont {T.}~\bibnamefont
  {Yoshimura}},\ }\href {\doibase 10.21468/SciPostPhys.2.2.014} {\bibfield
  {journal} {\bibinfo  {journal} {SciPost Phys.}\ }\textbf {\bibinfo {volume}
  {2}},\ \bibinfo {pages} {014} (\bibinfo {year} {2017})}\BibitemShut {NoStop}%
\bibitem [{\citenamefont {Ilievski}\ and\ \citenamefont
  {De~Nardis}(2017{\natexlab{a}})}]{PhysRevLett.119.020602}%
  \BibitemOpen
  \bibfield  {author} {\bibinfo {author} {\bibfnamefont {E.}~\bibnamefont
  {Ilievski}}\ and\ \bibinfo {author} {\bibfnamefont {J.}~\bibnamefont
  {De~Nardis}},\ }\href {\doibase 10.1103/PhysRevLett.119.020602} {\bibfield
  {journal} {\bibinfo  {journal} {Phys. Rev. Lett.}\ }\textbf {\bibinfo
  {volume} {119}},\ \bibinfo {pages} {020602} (\bibinfo {year}
  {2017}{\natexlab{a}})}\BibitemShut {NoStop}%
\bibitem [{\citenamefont {Bulchandani}\ \emph {et~al.}(2017)\citenamefont
  {Bulchandani}, \citenamefont {Vasseur}, \citenamefont {Karrasch},\ and\
  \citenamefont {Moore}}]{BBH0}%
  \BibitemOpen
  \bibfield  {author} {\bibinfo {author} {\bibfnamefont {V.~B.}\ \bibnamefont
  {Bulchandani}}, \bibinfo {author} {\bibfnamefont {R.}~\bibnamefont
  {Vasseur}}, \bibinfo {author} {\bibfnamefont {C.}~\bibnamefont {Karrasch}}, \
  and\ \bibinfo {author} {\bibfnamefont {J.~E.}\ \bibnamefont {Moore}},\ }\href
  {\doibase 10.1103/PhysRevLett.119.220604} {\bibfield  {journal} {\bibinfo
  {journal} {Phys. Rev. Lett.}\ }\textbf {\bibinfo {volume} {119}},\ \bibinfo
  {pages} {220604} (\bibinfo {year} {2017})}\BibitemShut {NoStop}%
\bibitem [{\citenamefont {Bulchandani}\ \emph {et~al.}(2018)\citenamefont
  {Bulchandani}, \citenamefont {Vasseur}, \citenamefont {Karrasch},\ and\
  \citenamefont {Moore}}]{BBH}%
  \BibitemOpen
  \bibfield  {author} {\bibinfo {author} {\bibfnamefont {V.~B.}\ \bibnamefont
  {Bulchandani}}, \bibinfo {author} {\bibfnamefont {R.}~\bibnamefont
  {Vasseur}}, \bibinfo {author} {\bibfnamefont {C.}~\bibnamefont {Karrasch}}, \
  and\ \bibinfo {author} {\bibfnamefont {J.~E.}\ \bibnamefont {Moore}},\ }\href
  {\doibase 10.1103/PhysRevB.97.045407} {\bibfield  {journal} {\bibinfo
  {journal} {Phys. Rev. B}\ }\textbf {\bibinfo {volume} {97}},\ \bibinfo
  {pages} {045407} (\bibinfo {year} {2018})}\BibitemShut {NoStop}%
\bibitem [{\citenamefont {Doyon}\ and\ \citenamefont
  {Spohn}(2017{\natexlab{a}})}]{GHDII}%
  \BibitemOpen
  \bibfield  {author} {\bibinfo {author} {\bibfnamefont {B.}~\bibnamefont
  {Doyon}}\ and\ \bibinfo {author} {\bibfnamefont {H.}~\bibnamefont {Spohn}},\
  }\href {\doibase 10.21468/SciPostPhys.3.6.039} {\bibfield  {journal}
  {\bibinfo  {journal} {SciPost Phys.}\ }\textbf {\bibinfo {volume} {3}},\
  \bibinfo {pages} {039} (\bibinfo {year} {2017}{\natexlab{a}})}\BibitemShut
  {NoStop}%
\bibitem [{\citenamefont {Doyon}\ and\ \citenamefont
  {Spohn}(2017{\natexlab{b}})}]{doyon2017dynamics}%
  \BibitemOpen
  \bibfield  {author} {\bibinfo {author} {\bibfnamefont {B.}~\bibnamefont
  {Doyon}}\ and\ \bibinfo {author} {\bibfnamefont {H.}~\bibnamefont {Spohn}},\
  }\href@noop {} {\bibfield  {journal} {\bibinfo  {journal} {Journal of
  Statistical Mechanics: Theory and Experiment}\ }\textbf {\bibinfo {volume}
  {2017}},\ \bibinfo {pages} {073210} (\bibinfo {year}
  {2017}{\natexlab{b}})}\BibitemShut {NoStop}%
\bibitem [{\citenamefont {Doyon}\ \emph {et~al.}(2018)\citenamefont {Doyon},
  \citenamefont {Yoshimura},\ and\ \citenamefont {Caux}}]{solitongases}%
  \BibitemOpen
  \bibfield  {author} {\bibinfo {author} {\bibfnamefont {B.}~\bibnamefont
  {Doyon}}, \bibinfo {author} {\bibfnamefont {T.}~\bibnamefont {Yoshimura}}, \
  and\ \bibinfo {author} {\bibfnamefont {J.-S.}\ \bibnamefont {Caux}},\ }\href
  {\doibase 10.1103/PhysRevLett.120.045301} {\bibfield  {journal} {\bibinfo
  {journal} {Phys. Rev. Lett.}\ }\textbf {\bibinfo {volume} {120}},\ \bibinfo
  {pages} {045301} (\bibinfo {year} {2018})}\BibitemShut {NoStop}%
\bibitem [{\citenamefont {Doyon}\ \emph {et~al.}(2017)\citenamefont {Doyon},
  \citenamefont {Dubail}, \citenamefont {Konik},\ and\ \citenamefont
  {Yoshimura}}]{PhysRevLett.119.195301}%
  \BibitemOpen
  \bibfield  {author} {\bibinfo {author} {\bibfnamefont {B.}~\bibnamefont
  {Doyon}}, \bibinfo {author} {\bibfnamefont {J.}~\bibnamefont {Dubail}},
  \bibinfo {author} {\bibfnamefont {R.}~\bibnamefont {Konik}}, \ and\ \bibinfo
  {author} {\bibfnamefont {T.}~\bibnamefont {Yoshimura}},\ }\href {\doibase
  10.1103/PhysRevLett.119.195301} {\bibfield  {journal} {\bibinfo  {journal}
  {Phys. Rev. Lett.}\ }\textbf {\bibinfo {volume} {119}},\ \bibinfo {pages}
  {195301} (\bibinfo {year} {2017})}\BibitemShut {NoStop}%
\bibitem [{\citenamefont {Collura}\ \emph {et~al.}(2018)\citenamefont
  {Collura}, \citenamefont {De~Luca},\ and\ \citenamefont
  {Viti}}]{PhysRevB.97.081111}%
  \BibitemOpen
  \bibfield  {author} {\bibinfo {author} {\bibfnamefont {M.}~\bibnamefont
  {Collura}}, \bibinfo {author} {\bibfnamefont {A.}~\bibnamefont {De~Luca}}, \
  and\ \bibinfo {author} {\bibfnamefont {J.}~\bibnamefont {Viti}},\ }\href
  {\doibase 10.1103/PhysRevB.97.081111} {\bibfield  {journal} {\bibinfo
  {journal} {Phys. Rev. B}\ }\textbf {\bibinfo {volume} {97}},\ \bibinfo
  {pages} {081111} (\bibinfo {year} {2018})}\BibitemShut {NoStop}%
\bibitem [{\citenamefont {De~Nardis}\ \emph
  {et~al.}(2018{\natexlab{a}})\citenamefont {De~Nardis}, \citenamefont
  {Bernard},\ and\ \citenamefont {Doyon}}]{dbd1}%
  \BibitemOpen
  \bibfield  {author} {\bibinfo {author} {\bibfnamefont {J.}~\bibnamefont
  {De~Nardis}}, \bibinfo {author} {\bibfnamefont {D.}~\bibnamefont {Bernard}},
  \ and\ \bibinfo {author} {\bibfnamefont {B.}~\bibnamefont {Doyon}},\ }\href
  {\doibase 10.1103/PhysRevLett.121.160603} {\bibfield  {journal} {\bibinfo
  {journal} {Phys. Rev. Lett.}\ }\textbf {\bibinfo {volume} {121}},\ \bibinfo
  {pages} {160603} (\bibinfo {year} {2018}{\natexlab{a}})}\BibitemShut
  {NoStop}%
\bibitem [{\citenamefont {Gopalakrishnan}\ \emph {et~al.}(2018)\citenamefont
  {Gopalakrishnan}, \citenamefont {Huse}, \citenamefont {Khemani},\ and\
  \citenamefont {Vasseur}}]{ghkv}%
  \BibitemOpen
  \bibfield  {author} {\bibinfo {author} {\bibfnamefont {S.}~\bibnamefont
  {Gopalakrishnan}}, \bibinfo {author} {\bibfnamefont {D.~A.}\ \bibnamefont
  {Huse}}, \bibinfo {author} {\bibfnamefont {V.}~\bibnamefont {Khemani}}, \
  and\ \bibinfo {author} {\bibfnamefont {R.}~\bibnamefont {Vasseur}},\
  }\href@noop {} {\bibfield  {journal} {\bibinfo  {journal} {arXiv preprint
  arXiv:1809.02126}\ } (\bibinfo {year} {2018})}\BibitemShut {NoStop}%
\bibitem [{\citenamefont {Ilievski}\ and\ \citenamefont
  {De~Nardis}(2017{\natexlab{b}})}]{PhysRevB.96.081118}%
  \BibitemOpen
  \bibfield  {author} {\bibinfo {author} {\bibfnamefont {E.}~\bibnamefont
  {Ilievski}}\ and\ \bibinfo {author} {\bibfnamefont {J.}~\bibnamefont
  {De~Nardis}},\ }\href {\doibase 10.1103/PhysRevB.96.081118} {\bibfield
  {journal} {\bibinfo  {journal} {Phys. Rev. B}\ }\textbf {\bibinfo {volume}
  {96}},\ \bibinfo {pages} {081118} (\bibinfo {year}
  {2017}{\natexlab{b}})}\BibitemShut {NoStop}%
\bibitem [{\citenamefont {Lebowitz}\ and\ \citenamefont
  {Percus}(1967)}]{percus}%
  \BibitemOpen
  \bibfield  {author} {\bibinfo {author} {\bibfnamefont {J.~L.}\ \bibnamefont
  {Lebowitz}}\ and\ \bibinfo {author} {\bibfnamefont {J.~K.}\ \bibnamefont
  {Percus}},\ }\href {\doibase 10.1103/PhysRev.155.122} {\bibfield  {journal}
  {\bibinfo  {journal} {Phys. Rev.}\ }\textbf {\bibinfo {volume} {155}},\
  \bibinfo {pages} {122} (\bibinfo {year} {1967})}\BibitemShut {NoStop}%
\bibitem [{\citenamefont {Sachdev}\ and\ \citenamefont
  {Young}(1997)}]{sachdev_young}%
  \BibitemOpen
  \bibfield  {author} {\bibinfo {author} {\bibfnamefont {S.}~\bibnamefont
  {Sachdev}}\ and\ \bibinfo {author} {\bibfnamefont {A.~P.}\ \bibnamefont
  {Young}},\ }\href {\doibase 10.1103/PhysRevLett.78.2220} {\bibfield
  {journal} {\bibinfo  {journal} {Phys. Rev. Lett.}\ }\textbf {\bibinfo
  {volume} {78}},\ \bibinfo {pages} {2220} (\bibinfo {year}
  {1997})}\BibitemShut {NoStop}%
\bibitem [{\citenamefont {Sachdev}\ and\ \citenamefont
  {Damle}(1997)}]{PhysRevLett.78.943}%
  \BibitemOpen
  \bibfield  {author} {\bibinfo {author} {\bibfnamefont {S.}~\bibnamefont
  {Sachdev}}\ and\ \bibinfo {author} {\bibfnamefont {K.}~\bibnamefont
  {Damle}},\ }\href {\doibase 10.1103/PhysRevLett.78.943} {\bibfield  {journal}
  {\bibinfo  {journal} {Phys. Rev. Lett.}\ }\textbf {\bibinfo {volume} {78}},\
  \bibinfo {pages} {943} (\bibinfo {year} {1997})}\BibitemShut {NoStop}%
\bibitem [{\citenamefont {Damle}\ and\ \citenamefont
  {Sachdev}(1998)}]{PhysRevB.57.8307}%
  \BibitemOpen
  \bibfield  {author} {\bibinfo {author} {\bibfnamefont {K.}~\bibnamefont
  {Damle}}\ and\ \bibinfo {author} {\bibfnamefont {S.}~\bibnamefont
  {Sachdev}},\ }\href {\doibase 10.1103/PhysRevB.57.8307} {\bibfield  {journal}
  {\bibinfo  {journal} {Phys. Rev. B}\ }\textbf {\bibinfo {volume} {57}},\
  \bibinfo {pages} {8307} (\bibinfo {year} {1998})}\BibitemShut {NoStop}%
\bibitem [{\citenamefont {El}(2003)}]{el2003thermodynamic}%
  \BibitemOpen
  \bibfield  {author} {\bibinfo {author} {\bibfnamefont {G.}~\bibnamefont
  {El}},\ }\href@noop {} {\bibfield  {journal} {\bibinfo  {journal} {Physics
  Letters A}\ }\textbf {\bibinfo {volume} {311}},\ \bibinfo {pages} {374}
  (\bibinfo {year} {2003})}\BibitemShut {NoStop}%
\bibitem [{\citenamefont {El}\ and\ \citenamefont {Kamchatnov}(2005)}]{el2005}%
  \BibitemOpen
  \bibfield  {author} {\bibinfo {author} {\bibfnamefont {G.~A.}\ \bibnamefont
  {El}}\ and\ \bibinfo {author} {\bibfnamefont {A.~M.}\ \bibnamefont
  {Kamchatnov}},\ }\href {\doibase 10.1103/PhysRevLett.95.204101} {\bibfield
  {journal} {\bibinfo  {journal} {Phys. Rev. Lett.}\ }\textbf {\bibinfo
  {volume} {95}},\ \bibinfo {pages} {204101} (\bibinfo {year}
  {2005})}\BibitemShut {NoStop}%
\bibitem [{\citenamefont {Damle}\ and\ \citenamefont
  {Sachdev}(2005)}]{PhysRevLett.95.187201}%
  \BibitemOpen
  \bibfield  {author} {\bibinfo {author} {\bibfnamefont {K.}~\bibnamefont
  {Damle}}\ and\ \bibinfo {author} {\bibfnamefont {S.}~\bibnamefont
  {Sachdev}},\ }\href {\doibase 10.1103/PhysRevLett.95.187201} {\bibfield
  {journal} {\bibinfo  {journal} {Phys. Rev. Lett.}\ }\textbf {\bibinfo
  {volume} {95}},\ \bibinfo {pages} {187201} (\bibinfo {year}
  {2005})}\BibitemShut {NoStop}%
\bibitem [{\citenamefont {Sirker}\ \emph {et~al.}(2009)\citenamefont {Sirker},
  \citenamefont {Pereira},\ and\ \citenamefont {Affleck}}]{sirker:2010}%
  \BibitemOpen
  \bibfield  {author} {\bibinfo {author} {\bibfnamefont {J.}~\bibnamefont
  {Sirker}}, \bibinfo {author} {\bibfnamefont {R.~G.}\ \bibnamefont {Pereira}},
  \ and\ \bibinfo {author} {\bibfnamefont {I.}~\bibnamefont {Affleck}},\ }\href
  {\doibase 10.1103/PhysRevLett.103.216602} {\bibfield  {journal} {\bibinfo
  {journal} {Phys. Rev. Lett.}\ }\textbf {\bibinfo {volume} {103}},\ \bibinfo
  {pages} {216602} (\bibinfo {year} {2009})}\BibitemShut {NoStop}%
\bibitem [{\citenamefont {Sirker}\ \emph {et~al.}(2011)\citenamefont {Sirker},
  \citenamefont {Pereira},\ and\ \citenamefont {Affleck}}]{PhysRevB.83.035115}%
  \BibitemOpen
  \bibfield  {author} {\bibinfo {author} {\bibfnamefont {J.}~\bibnamefont
  {Sirker}}, \bibinfo {author} {\bibfnamefont {R.~G.}\ \bibnamefont {Pereira}},
  \ and\ \bibinfo {author} {\bibfnamefont {I.}~\bibnamefont {Affleck}},\ }\href
  {\doibase 10.1103/PhysRevB.83.035115} {\bibfield  {journal} {\bibinfo
  {journal} {Phys. Rev. B}\ }\textbf {\bibinfo {volume} {83}},\ \bibinfo
  {pages} {035115} (\bibinfo {year} {2011})}\BibitemShut {NoStop}%
\bibitem [{\citenamefont {Medenjak}\ \emph
  {et~al.}(2017{\natexlab{a}})\citenamefont {Medenjak}, \citenamefont
  {Klobas},\ and\ \citenamefont {Prosen}}]{medenjak2017}%
  \BibitemOpen
  \bibfield  {author} {\bibinfo {author} {\bibfnamefont {M.}~\bibnamefont
  {Medenjak}}, \bibinfo {author} {\bibfnamefont {K.}~\bibnamefont {Klobas}}, \
  and\ \bibinfo {author} {\bibfnamefont {T.~c.~v.}\ \bibnamefont {Prosen}},\
  }\href {\doibase 10.1103/PhysRevLett.119.110603} {\bibfield  {journal}
  {\bibinfo  {journal} {Phys. Rev. Lett.}\ }\textbf {\bibinfo {volume} {119}},\
  \bibinfo {pages} {110603} (\bibinfo {year} {2017}{\natexlab{a}})}\BibitemShut
  {NoStop}%
\bibitem [{\citenamefont {Kormos}\ \emph {et~al.}(2018)\citenamefont {Kormos},
  \citenamefont {Moca},\ and\ \citenamefont {Zar\'and}}]{kormos2017}%
  \BibitemOpen
  \bibfield  {author} {\bibinfo {author} {\bibfnamefont {M.}~\bibnamefont
  {Kormos}}, \bibinfo {author} {\bibfnamefont {C.~P.}\ \bibnamefont {Moca}}, \
  and\ \bibinfo {author} {\bibfnamefont {G.}~\bibnamefont {Zar\'and}},\ }\href
  {\doibase 10.1103/PhysRevE.98.032105} {\bibfield  {journal} {\bibinfo
  {journal} {Phys. Rev. E}\ }\textbf {\bibinfo {volume} {98}},\ \bibinfo
  {pages} {032105} (\bibinfo {year} {2018})}\BibitemShut {NoStop}%
\bibitem [{\citenamefont {Piroli}\ \emph {et~al.}(2017)\citenamefont {Piroli},
  \citenamefont {De~Nardis}, \citenamefont {Collura}, \citenamefont {Bertini},\
  and\ \citenamefont {Fagotti}}]{piroli2017}%
  \BibitemOpen
  \bibfield  {author} {\bibinfo {author} {\bibfnamefont {L.}~\bibnamefont
  {Piroli}}, \bibinfo {author} {\bibfnamefont {J.}~\bibnamefont {De~Nardis}},
  \bibinfo {author} {\bibfnamefont {M.}~\bibnamefont {Collura}}, \bibinfo
  {author} {\bibfnamefont {B.}~\bibnamefont {Bertini}}, \ and\ \bibinfo
  {author} {\bibfnamefont {M.}~\bibnamefont {Fagotti}},\ }\href {\doibase
  10.1103/PhysRevB.96.115124} {\bibfield  {journal} {\bibinfo  {journal} {Phys.
  Rev. B}\ }\textbf {\bibinfo {volume} {96}},\ \bibinfo {pages} {115124}
  (\bibinfo {year} {2017})}\BibitemShut {NoStop}%
\bibitem [{\citenamefont {Klobas}\ \emph {et~al.}(2018)\citenamefont {Klobas},
  \citenamefont {Medenjak}, \citenamefont {Prosen},\ and\ \citenamefont
  {Vanicat}}]{klobas2018a}%
  \BibitemOpen
  \bibfield  {author} {\bibinfo {author} {\bibfnamefont {K.}~\bibnamefont
  {Klobas}}, \bibinfo {author} {\bibfnamefont {M.}~\bibnamefont {Medenjak}},
  \bibinfo {author} {\bibfnamefont {T.}~\bibnamefont {Prosen}}, \ and\ \bibinfo
  {author} {\bibfnamefont {M.}~\bibnamefont {Vanicat}},\ }\href@noop {}
  {\bibfield  {journal} {\bibinfo  {journal} {arXiv preprint arXiv:1807.05000}\
  } (\bibinfo {year} {2018})}\BibitemShut {NoStop}%
\bibitem [{\citenamefont {Klobas}\ \emph {et~al.}()\citenamefont {Klobas},
  \citenamefont {Medenjak},\ and\ \citenamefont {Prosen}}]{klobas2018b}%
  \BibitemOpen
  \bibfield  {author} {\bibinfo {author} {\bibfnamefont {K.}~\bibnamefont
  {Klobas}}, \bibinfo {author} {\bibfnamefont {M.}~\bibnamefont {Medenjak}}, \
  and\ \bibinfo {author} {\bibfnamefont {T.}~\bibnamefont {Prosen}},\
  }\href@noop {} {\bibinfo  {journal} {arXiv:1808.07385}\ }\BibitemShut
  {NoStop}%
\bibitem [{\citenamefont {Gopalakrishnan}(2018)}]{sg_ffa}%
  \BibitemOpen
\bibfield  {journal} {  }\bibfield  {author} {\bibinfo {author} {\bibfnamefont
  {S.}~\bibnamefont {Gopalakrishnan}},\ }\href {\doibase
  10.1103/PhysRevB.98.060302} {\bibfield  {journal} {\bibinfo  {journal} {Phys.
  Rev. B}\ }\textbf {\bibinfo {volume} {98}},\ \bibinfo {pages} {060302}
  (\bibinfo {year} {2018})}\BibitemShut {NoStop}%
\bibitem [{Note1()}]{Note1}%
  \BibitemOpen
  \bibinfo {note} {We are concerned here with finite-temperature states but
  \protect \emph {not} with generalized Gibbs ensembles with finite
  magnetization. In the finite-magnetization sectors, spin transport remains
  ballistic for all $\Delta $.}\BibitemShut {Stop}%
\bibitem [{\citenamefont {Karrasch}\ \emph {et~al.}(2014)\citenamefont
  {Karrasch}, \citenamefont {Moore},\ and\ \citenamefont
  {Heidrich-Meisner}}]{PhysRevB.89.075139}%
  \BibitemOpen
  \bibfield  {author} {\bibinfo {author} {\bibfnamefont {C.}~\bibnamefont
  {Karrasch}}, \bibinfo {author} {\bibfnamefont {J.~E.}\ \bibnamefont {Moore}},
  \ and\ \bibinfo {author} {\bibfnamefont {F.}~\bibnamefont
  {Heidrich-Meisner}},\ }\href {\doibase 10.1103/PhysRevB.89.075139} {\bibfield
   {journal} {\bibinfo  {journal} {Phys. Rev. B}\ }\textbf {\bibinfo {volume}
  {89}},\ \bibinfo {pages} {075139} (\bibinfo {year} {2014})}\BibitemShut
  {NoStop}%
\bibitem [{\citenamefont {Karrasch}(2017)}]{1367-2630-19-3-033027}%
  \BibitemOpen
  \bibfield  {author} {\bibinfo {author} {\bibfnamefont {C.}~\bibnamefont
  {Karrasch}},\ }\href {http://stacks.iop.org/1367-2630/19/i=3/a=033027}
  {\bibfield  {journal} {\bibinfo  {journal} {New Journal of Physics}\ }\textbf
  {\bibinfo {volume} {19}},\ \bibinfo {pages} {033027} (\bibinfo {year}
  {2017})}\BibitemShut {NoStop}%
\bibitem [{\citenamefont {Bouchaud}\ and\ \citenamefont
  {Georges}(1990)}]{bouchaud1990}%
  \BibitemOpen
  \bibfield  {author} {\bibinfo {author} {\bibfnamefont {J.-P.}\ \bibnamefont
  {Bouchaud}}\ and\ \bibinfo {author} {\bibfnamefont {A.}~\bibnamefont
  {Georges}},\ }\href@noop {} {\bibfield  {journal} {\bibinfo  {journal}
  {Physics reports}\ }\textbf {\bibinfo {volume} {195}},\ \bibinfo {pages}
  {127} (\bibinfo {year} {1990})}\BibitemShut {NoStop}%
\bibitem [{\citenamefont {Takahashi}(1999)}]{Takahashi}%
  \BibitemOpen
  \bibfield  {author} {\bibinfo {author} {\bibfnamefont {M.}~\bibnamefont
  {Takahashi}},\ }\href {https://books.google.com/books?id=kX1FAwEACAAJ} {\emph
  {\bibinfo {title} {Thermodynamics of One-Dimensional Solvable Models}}}\
  (\bibinfo  {publisher} {Cambridge University Press},\ \bibinfo {year}
  {1999})\BibitemShut {NoStop}%
\bibitem [{\citenamefont {Ganahl}\ \emph {et~al.}(2013)\citenamefont {Ganahl},
  \citenamefont {Haque},\ and\ \citenamefont {Evertz}}]{quantumbowling}%
  \BibitemOpen
  \bibfield  {author} {\bibinfo {author} {\bibfnamefont {M.}~\bibnamefont
  {Ganahl}}, \bibinfo {author} {\bibfnamefont {M.}~\bibnamefont {Haque}}, \
  and\ \bibinfo {author} {\bibfnamefont {H.}~\bibnamefont {Evertz}},\
  }\href@noop {} {\bibfield  {journal} {\bibinfo  {journal} {arXiv preprint
  arXiv:1302.2667}\ } (\bibinfo {year} {2013})}\BibitemShut {NoStop}%
\bibitem [{\citenamefont {Medenjak}\ \emph
  {et~al.}(2017{\natexlab{b}})\citenamefont {Medenjak}, \citenamefont
  {Karrasch},\ and\ \citenamefont {Prosen}}]{mkp}%
  \BibitemOpen
  \bibfield  {author} {\bibinfo {author} {\bibfnamefont {M.}~\bibnamefont
  {Medenjak}}, \bibinfo {author} {\bibfnamefont {C.}~\bibnamefont {Karrasch}},
  \ and\ \bibinfo {author} {\bibfnamefont {T.~c.~v.}\ \bibnamefont {Prosen}},\
  }\href {\doibase 10.1103/PhysRevLett.119.080602} {\bibfield  {journal}
  {\bibinfo  {journal} {Phys. Rev. Lett.}\ }\textbf {\bibinfo {volume} {119}},\
  \bibinfo {pages} {080602} (\bibinfo {year} {2017}{\natexlab{b}})}\BibitemShut
  {NoStop}%
\bibitem [{\citenamefont {Bonnes}\ \emph {et~al.}(2014)\citenamefont {Bonnes},
  \citenamefont {Essler},\ and\ \citenamefont
  {L\"auchli}}]{PhysRevLett.113.187203}%
  \BibitemOpen
  \bibfield  {author} {\bibinfo {author} {\bibfnamefont {L.}~\bibnamefont
  {Bonnes}}, \bibinfo {author} {\bibfnamefont {F.~H.~L.}\ \bibnamefont
  {Essler}}, \ and\ \bibinfo {author} {\bibfnamefont {A.~M.}\ \bibnamefont
  {L\"auchli}},\ }\href {\doibase 10.1103/PhysRevLett.113.187203} {\bibfield
  {journal} {\bibinfo  {journal} {Phys. Rev. Lett.}\ }\textbf {\bibinfo
  {volume} {113}},\ \bibinfo {pages} {187203} (\bibinfo {year}
  {2014})}\BibitemShut {NoStop}%
\bibitem [{\citenamefont {Lieb}\ and\ \citenamefont
  {Robinson}(2004)}]{Lieb2004}%
  \BibitemOpen
  \bibfield  {author} {\bibinfo {author} {\bibfnamefont {E.~H.}\ \bibnamefont
  {Lieb}}\ and\ \bibinfo {author} {\bibfnamefont {D.~W.}\ \bibnamefont
  {Robinson}},\ }\enquote {\bibinfo {title} {The finite group velocity of
  quantum spin systems},}\ in\ \href {\doibase 10.1007/978-3-662-10018-9_25}
  {\emph {\bibinfo {booktitle} {Statistical Mechanics: Selecta of Elliott H.
  Lieb}}},\ \bibinfo {editor} {edited by\ \bibinfo {editor} {\bibfnamefont
  {B.}~\bibnamefont {Nachtergaele}}, \bibinfo {editor} {\bibfnamefont {J.~P.}\
  \bibnamefont {Solovej}}, \ and\ \bibinfo {editor} {\bibfnamefont
  {J.}~\bibnamefont {Yngvason}}}\ (\bibinfo  {publisher} {Springer Berlin
  Heidelberg},\ \bibinfo {address} {Berlin, Heidelberg},\ \bibinfo {year}
  {2004})\ pp.\ \bibinfo {pages} {425--431}\BibitemShut {NoStop}%
\bibitem [{\citenamefont {Fukuhara}\ \emph {et~al.}(2013)\citenamefont
  {Fukuhara}, \citenamefont {Schau{\ss}}, \citenamefont {Endres}, \citenamefont
  {Hild}, \citenamefont {Cheneau}, \citenamefont {Bloch},\ and\ \citenamefont
  {Gross}}]{fukuhara2013}%
  \BibitemOpen
  \bibfield  {author} {\bibinfo {author} {\bibfnamefont {T.}~\bibnamefont
  {Fukuhara}}, \bibinfo {author} {\bibfnamefont {P.}~\bibnamefont
  {Schau{\ss}}}, \bibinfo {author} {\bibfnamefont {M.}~\bibnamefont {Endres}},
  \bibinfo {author} {\bibfnamefont {S.}~\bibnamefont {Hild}}, \bibinfo {author}
  {\bibfnamefont {M.}~\bibnamefont {Cheneau}}, \bibinfo {author} {\bibfnamefont
  {I.}~\bibnamefont {Bloch}}, \ and\ \bibinfo {author} {\bibfnamefont
  {C.}~\bibnamefont {Gross}},\ }\href@noop {} {\bibfield  {journal} {\bibinfo
  {journal} {Nature}\ }\textbf {\bibinfo {volume} {502}},\ \bibinfo {pages}
  {76} (\bibinfo {year} {2013})}\BibitemShut {NoStop}%
\bibitem [{\citenamefont {St{\'e}phan}(2017)}]{1742-5468-2017-10-103108}%
  \BibitemOpen
  \bibfield  {author} {\bibinfo {author} {\bibfnamefont {J.-M.}\ \bibnamefont
  {St{\'e}phan}},\ }\href {http://stacks.iop.org/1742-5468/2017/i=10/a=103108}
  {\bibfield  {journal} {\bibinfo  {journal} {Journal of Statistical Mechanics:
  Theory and Experiment}\ }\textbf {\bibinfo {volume} {2017}},\ \bibinfo
  {pages} {103108} (\bibinfo {year} {2017})}\BibitemShut {NoStop}%
\bibitem [{\citenamefont {Misguich}\ \emph {et~al.}(2017)\citenamefont
  {Misguich}, \citenamefont {Mallick},\ and\ \citenamefont
  {Krapivsky}}]{PhysRevB.96.195151}%
  \BibitemOpen
  \bibfield  {author} {\bibinfo {author} {\bibfnamefont {G.}~\bibnamefont
  {Misguich}}, \bibinfo {author} {\bibfnamefont {K.}~\bibnamefont {Mallick}}, \
  and\ \bibinfo {author} {\bibfnamefont {P.~L.}\ \bibnamefont {Krapivsky}},\
  }\href {\doibase 10.1103/PhysRevB.96.195151} {\bibfield  {journal} {\bibinfo
  {journal} {Phys. Rev. B}\ }\textbf {\bibinfo {volume} {96}},\ \bibinfo
  {pages} {195151} (\bibinfo {year} {2017})}\BibitemShut {NoStop}%
\bibitem [{\citenamefont {{Bulchandani}}\ and\ \citenamefont
  {{Karrasch}}(2018)}]{2018arXiv181008227B}%
  \BibitemOpen
  \bibfield  {author} {\bibinfo {author} {\bibfnamefont {V.~B.}\ \bibnamefont
  {{Bulchandani}}}\ and\ \bibinfo {author} {\bibfnamefont {C.}~\bibnamefont
  {{Karrasch}}},\ }\href@noop {} {\bibfield  {journal} {\bibinfo  {journal}
  {ArXiv e-prints}\ ,\ \bibinfo {eid} {arXiv:1810.08227}} (\bibinfo {year}
  {2018})},\ \Eprint {http://arxiv.org/abs/1810.08227} {arXiv:1810.08227
  [cond-mat.stat-mech]} \BibitemShut {NoStop}%
\bibitem [{Note2()}]{Note2}%
  \BibitemOpen
  \bibinfo {note} {S. Gopalakrishnan and V. Alba, in preparation.}\BibitemShut
  {Stop}%
\bibitem [{\citenamefont {De~Nardis}\ \emph
  {et~al.}(2018{\natexlab{b}})\citenamefont {De~Nardis}, \citenamefont
  {Bernard},\ and\ \citenamefont {Doyon}}]{dbd2}%
  \BibitemOpen
  \bibfield  {author} {\bibinfo {author} {\bibfnamefont {J.}~\bibnamefont
  {De~Nardis}}, \bibinfo {author} {\bibfnamefont {D.}~\bibnamefont {Bernard}},
  \ and\ \bibinfo {author} {\bibfnamefont {B.}~\bibnamefont {Doyon}},\
  }\href@noop {} {\bibfield  {journal} {\bibinfo  {journal} {arXiv:1812.00767}\
  } (\bibinfo {year} {2018}{\natexlab{b}})}\BibitemShut {NoStop}%
\end{thebibliography}%

\end{document}